\documentclass[aps,prl,longbibliography,twocolumn,showpacs,noeprint]{revtex4-2}

\usepackage{CJK}

\usepackage{graphicx}

\usepackage{amsmath}
\usepackage{dcolumn}

\usepackage{bm}
\usepackage[normalem]{ulem}
\usepackage{color}
\usepackage{hyperref}

\usepackage{epstopdf}
\newcommand{\bew}{\begin{widetext}}
\newcommand{\ew}{\end{widetext}}

\newcommand{\ii}{{\rm i}}
\newcommand{\bx}{\mathbf{x}}

\newcommand{\bp}{\mathbf{p}}
\newcommand{\bq}{\mathbf{q}}
\newcommand{\bv}{\mathbf{v}}

\newcommand{\br}{\mathbf{r}}

\newcommand{\bff}{\mathbf{f}}
\newcommand{\bu}{\mathbf{u}}

\newcommand{\bk}{\mathbf{k}}

\newcommand{\sep}{ \ \ \ , \ \ \ }

\newcommand{\beq}{\begin{equation}}
\newcommand{\eeq}{\end{equation}}
\newcommand{\beqn}{\begin{eqnarray}}
\newcommand{\eeqn}{\end{eqnarray}}
\newcommand{\pp}{\partial}
\newcommand{\dd}{{\rm d}}

\newcommand{\fig}{Fig.\ }

\newcommand{\cP}{{\cal P}}

\newcommand{\la}{\langle}

\newcommand{\ra}{\rangle}

\newcommand{\vnab}{{\bf \nabla}}

\begin{document}
\title{Can exact scaling exponents be obtained using the renormalization group?\\
 Affirmative evidence from incompressible polar active fluids } 
\author{Patrick Jentsch}
\email{p.jentsch20@imperial.ac.uk}
\address{Department of Bioengineering, Imperial College London, South Kensington Campus, London SW7 2AZ, U.K.}
\author{Chiu Fan Lee}
\email{c.lee@imperial.ac.uk}
\address{Department of Bioengineering, Imperial College London, South Kensington Campus, London SW7 2AZ, U.K.}
\date{\today}

	\begin{abstract}
In active matter systems, non-Gaussian, exact scaling exponents have been claimed in a range of systems using perturbative renormalization group (RG) methods. This is unusual compared to equilibrium systems where non-Gaussian exponents can typically only be approximated, even using the exact (or functional/nonperturbative) renormalization group (ERG). Here, we perform an ERG analysis on the ordered phase of incompressible polar active fluids and find that the {\it exact} non-Gaussian exponents  obtained  previously using a perturbative RG method remain valid even in this nonperturbative setting. Furthermore, our ERG analysis elucidates the RG flow of this system and enables us to identify an active Goldstone regime with nontrivial, long-ranged scaling behavior for parallel and longitudinal fluctuations.
	\end{abstract}
	
\maketitle

Renormalization group (RG) methodology constituted one of the greatest advances in the toolbox of theoretical physicists in the past 50 years and has brought many great advances in physics since its inception. Originated from particle and condensed matter physics \cite{gell-mann_physrev54,symanzik_CMPhys70,callan_prd70,kadanoff_ppf66,wilson_prb71a,wilson_prb71b}, RG techniques have since found applications in diverse disciplines of physics. In the context of many-body physics, RG methods enable us to identify emergent behavior that is {\it universal} to a wide class of systems sharing the same key  qualitative  characteristics features, such as the underlying  conservation laws and symmetries \cite{hohenberg_rmp77,chaikin_b1995}. Furthermore, RG provides us with a way to classify many-body systems into distinct {\it universality classes} (UCs), each of which is associated with a unique RG fixed point.  Importantly, distinct UCs typically exhibit quantitatively different  scale-invariant structures and thus leave measurable experimental imprints. 

Interestingly, this also provides a  way to ascertain novelty in physics: a system can be said to exhibit novel physics if it is governed by a novel UC. 
In this regard, the nascent field of active matter, nonequilibrium many-body systems that generate local stresses at the constituent-level \cite{ramaswamy_annrev10,marchetti_rmp13}, has been a treasure trove of novel UCs: diverse new critical phenomena and nonequilibrium phases have been uncovered in the recent past (see \cite{chen_prl20,mahdisoltani_prr21, van_der_kolk_a2022,chen_prl22b,zinati_pre22,cavagna_natphys2023,jentsch_prr23,jentsch_a24, miller_a24} for recent examples).

However, while the novelty of these dynamical systems can typically be identified through analytical RG calculations, the accompanying quantitative features can be more difficult to discern. This is partly because RG calculations have historically been perturbative in nature, with the $\epsilon$-expansion method being one of the most popular methods used \cite{wilson_prl72,wilson_physrep74}. In an $\epsilon$-expansion, the supposed ``small" parameter $\epsilon$ corresponds to the value between the spatial dimension of interest and a model-dependent upper critical dimension, $d_u$. Unfortunately, $d_u$ is for many systems beyond any physical dimensions (e.g., $d_u = 4$ for the critical Ising model), thus making a quantitative RG calculation using the $\epsilon$-expansion method in physical dimensions, where $\epsilon=1$ or $\epsilon=2$, questionable. 

Undeterred, physicists continued to make great strides in developing RG methodology. In particular, tremendous advances have been made in exact (or 
functional/nonperturbative) RG methods \cite{wetterich_plb93,ellwanger_zfpc94,morris_ijmpa94}, which was shown to be quantitatively accurate when applied to diverse physical systems \cite{dupuis_pr21}. Despite the namesake, practitioners of exact RG (ERG) calculations almost never claim that their outputs, such as the scaling exponents computed, are actually {\it exact} when dealing with a nontrivial RG fixed point.
This is because an ERG calculation is invariably coupled to an approximation scheme, such as the derivative expansion \cite{berges_pr02,canet_prd03,balog_prl19,depolsi_pre20} or the BMW approximation \cite{blaizot_plb06,benitez_pre09,benitez_pre12}. 
The accuracy of scaling exponents obtained in such an approximation can typically be improved, by incorporating higher-order terms which are irrelevant by naive power-counting. For example, the convergence of the derivative expansion to the virtually exact exponents has been demonstrated quantitatively for the critical point of $O(N)$ models \cite{balog_prl19,depolsi_pre20}.

Since in general it is impossible to perform an ERG calculation on a completely generic Hamiltonian (i.e., with infinitely many terms), no exact results can be expected.  Ironically,  practitioners of the perturbative dynamic RG (DRG) \cite{forster_pra77} have long claimed that they have found numerically exact scaling exponents across a spectrum of dimensions in biology-inspired systems \cite{hwa_prl89, toner_prl95,chen_njp18,toner_prl18,mahdisoltani_prr21,chen_prl22}. So how can both observations be reconciled?

In this Letter, we provide strong evidence that for some systems exact calculations can be performed using RG methods.
Specifically, we apply ERG to analyze the ordered phase of incompressible polar active fluids (IPAF) in three dimensions, whose associate scaling exponents were claimed to be determined exactly using the perturbative DRG method \cite{chen_njp18}. 

By performing an ERG calculation on the same system from scratch, we confirm the existence of the fixed point, which previously was only assumed, and find that the scaling exponents \cite{chen_njp18} remain unchanged, thus affirming the {\it exact} nature of these quantities. 
Further, we find an {\it active Goldstone regime}, where two other modes: velocity fluctuations that are aligned with collective motion and wavevector respectively, become soft and exhibit nontrivial scaling behavior.

In the following, we will first recapitulate the key arguments in the DRG calculation in Ref.~\cite{chen_njp18} that lead to the claim of exact scaling exponents. 
We then reanalyze IPAF using out-of-equilibrium ERG \cite{canet_jopa11} 
with a more general ansatz, and show that the scaling exponents remain unmodified. In the course of the analysis, we will find a more general fixed point than described before, realizing the active Goldstone regime.

{\it A recap of DRG on IPAF.}---The equation of motion (EOM) that governs generic IPAF corresponds to the incompressible version of the Toner-Tu EOM for generic compressible polar active fluids. Specifically, denoting the system's velocity field by $\bv$, the EOM is 
\beq
\label{eq:eom}
\pp_t \bv +\lambda (\bv \cdot \vnab) \bv = -\vnab \cP -(a+b |\bv|^2)\bv +\mu \vnab^2 \bv + {\rm h.o.t.}+\bff 
 ,
\eeq
where $\cP$ is the ``pressure" term (or Lagrange multiplier) present to enforce the incompressibility condition $\vnab \cdot \bv=0$ and  ``h.o.t." denotes {\it higher order terms}, i.e., terms of higher order in both $\bv$ and the spatial derivatives. Finally, $\bff$ is  a zero-mean Gaussian noise with statistics:
\beq
\la f_m(\br, t) f_n(\br',t')\ra= 2D \delta^d (\br+\br')  \delta (t+t') .
\eeq

Since in the ordered phase the continuous rotational symmetry is broken spontaneously, we expect that the resulting Goldstone modes exhibit scaling behavior  that is described by a RG fixed point. 
Specifically, letting $\bu= \bv - |\la \bv \ra| \hat{\bf x}$ where $\hat{\bf x}$ denotes, without loss of generality,  the direction of the collective motion $\la \bv \ra$, we expect that 
\beq
\label{eq:S}
\langle\bu_\perp(\mathbf 0,0)\cdot\bu_\perp(\br,t)\rangle= |\br_\perp|^{2\chi}S\left(\frac{x- \nu t}{|\br_\perp|^\zeta}, \frac{t}{|\br_\perp|^z}\right)
\ ,
\eeq
where  ``$\perp$'' denotes components perpendicular to $\hat{\bf x}$ and so $\bu_\perp$ corresponds to the Goldstone modes in the ordered phase.  Furthermore, $S$ in Eq.~(\ref{eq:S}) is a scaling function that is universal up to a model-dependent constant prefactor, and $\nu$ is again a model-dependent constant.

Using a DRG analysis, it is claimed in Ref.~\cite{chen_njp18} that in $2<d\leq 4$, the values of the scaling exponents are {\it exactly} given by
\beq
\label{exact}
 \chi=\frac{3-2d}{5} \sep \zeta=\frac{d+1}{5}\sep z=\frac{2(d+1)}{5} \ .
\eeq

We now summarize the chain of arguments leading to the claim of exact scaling exponents that describe the ordered phase of IPAF.

Step 1.\ An analysis of the linearized version of the EOM (\ref{eq:eom}) indicates that the correlation function $\la \bu (\bk,t) \cdot \bu(\bk',t') \ra$ is dominated by $\la \bu_T (\bk,t) \cdot \bu_T(\bk',t') \ra$ where $\bu_T(\bk,t) \equiv \bu_\perp(\bk,t) - [\bu_\perp(\bk,t) \cdot \hat{\bk} ] \hat{\bk}$.

Step 2.\
After determining the dominant components in the fluctuations, the most dominant nonlinear terms in the EOM are identified by power counting. Retaining only the most relevant nonlinear term, the reduced EOM of $\bu_\perp$, in the comoving frame along $\hat{\bx}$, is found to be
\beqn
\label{eq:uperp}
\nonumber
\pp_t \bu_\perp +\lambda (\bu_\perp \cdot \vnab_\perp) \bu_\perp &=& -\vnab_\perp \cP +\mu_\perp \nabla_\perp^2 \bu_\perp
\\
&&+\mu_x \pp_x^2 \bu_\perp +\bff_\perp
\ .
\eeqn
In particular, the upper critical dimension $d_u$ is 4.

Step 3.\
The RG flow equations of the four model coefficients ($\lambda, \mu_\perp, \mu_x$ and $D$) evaluated at the fixed point (that is assumed to exist) lead to four linear algebraic equations in terms of the yet to be determined scaling exponents $\chi, \zeta$, and $z$, and potential graphical corrections. However, since the structure of the EOM corresponds exactly to the model equation analyzed by Toner and Tu in 1995 \cite{toner_prl95}, we know that only one of the coefficients ($\mu_\perp$) admits a graphical correction ($G_{\mu_\perp}$). The four linear equations obtained at the RG fixed point thus enable us to solve for the four unknowns: $\chi, \zeta, z$ and $G_{\mu_\perp}$, using simple linear algebra, yielding Eq.~\eqref{exact}.

Step 4.\
One can now use the scaling exponents obtained to check that all other nonlinear terms ignored in the analysis remain irrelevant for $d=3$. Therefore, the scaling behavior of the system is claimed to be described by the exact scaling exponents obtained.

\begin{figure*}[t]
\includegraphics[width=\linewidth]{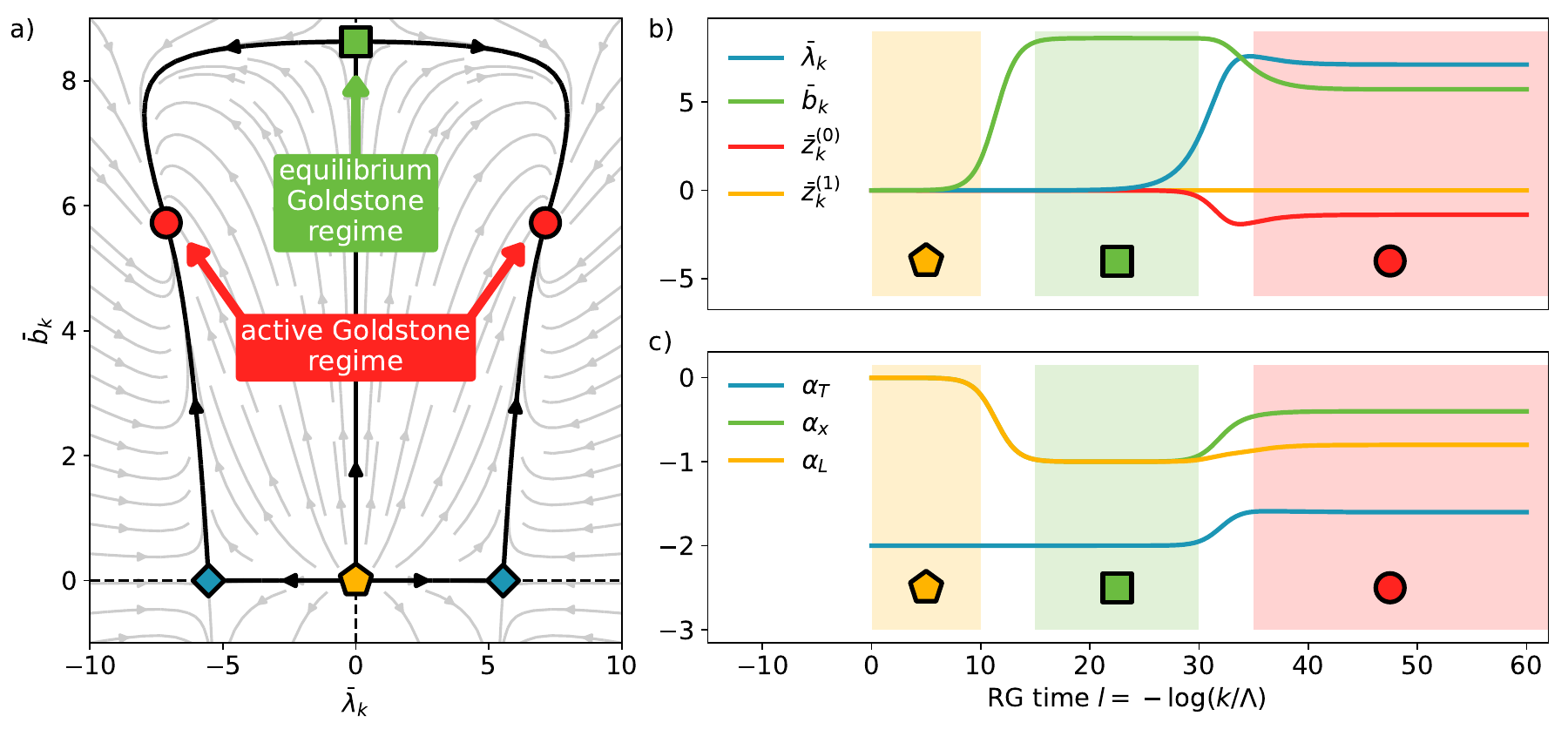}
\caption{
a) Two-dimensional projection of the RG flow diagram in $d=3$ (how the projection is obtained is explained in Ref.~\cite{SI}). The yellow pentagon denotes the trivial Gaussian fixed point and the blue diamond the universality class described in \cite{chen_njp18}. At the green square, the system is in the Goldstone regime of the equilibrium $O(N)$ model (for $N=d-1$). Finally the red circle denotes the active Goldstone regime described in this paper. b) A specific RG trajectory in $d=3$ which shows a crossover from the Gaussian fixed point (yellow pentagon) over the equilibrium Goldstone regime (green square) to the active Goldstone regime (red circle). c) The scaling dimension of the 3 different dynamical modes along the same trajectory as in b). In the active Goldstone regime the Goldstone modes scaling dimension, $\alpha_T$, agrees with the value calculated in Ref.~\cite{chen_njp18}, while the other two modes not considered in Ref.~\cite{chen_njp18}, $\alpha_T$ and $\alpha_L$, show novel scaling behavior. 
}
\label{fig:flow}
\end{figure*}
{\it ERG on IPAF.---}We now reanalyze the ordered phase of IPAF from scratch to answer the questions: Does the fixed point actually exist? And can nonperturbative effects modify the scaling behavior \eqref{exact}? Akin to the treatment of passive incompressible fluids with long-ranged forcing, described by the Navier-Stokes equation \cite{tomassini_plb97,canet_pre15}, we first convert the EOM \eqref{eq:eom} to an action using the Martin-Siggia-Rose-De Dominicis-Janssen formalism  \cite{martin_pra73,dedominics_JPhCol76,janssen_ZPhB76}, keeping the pressure as an auxiliary variable that enforces the incompressibility condition,
\begin{align}
\label{eq:microaction}
\nonumber
S&\left[\bar\bv,\bv,\bar\cP, \cP\right] = \int_{\tilde \br}\bigg\{ \bar\bv\cdot \bigg[ \pp_t \bv +\lambda (\bv \cdot \vnab) \bv + \vnab \cP-\mu \vnab^2 \bv   \\
 & +(a+b |\bv|^2)\bv \bigg] - D|\bar\bv|^2 + \bar\cP \nabla\cdot {\bv}\bigg\}
 ,
\end{align}
where $\int_{\tilde \br}=\int\dd^d \br\dd t$ and $\bar\bv$ and $\bar\cP$ are the response fields introduced by the formalism.

Expressed in this form, the functional renormalization group formalism, based on the exact Wetterich equation \cite{wetterich_plb93,ellwanger_zfpc94,morris_ijmpa94},
\beq
\label{eq:wetterich}
\pp_k \Gamma_k =\frac{1}{2} {\rm Tr} \left[ \left(\Gamma^{(2)}_k +R_k\right)^{-1} \pp_k R_k\right]\ ,
\eeq
where the trace ${\rm Tr}$ sums over all degrees of freedom, i.e., field indices, wavenumbers, and frequencies, can now straightforwardly be applied. Eq.~\eqref{eq:wetterich} describes the coarse-graining flow of $\Gamma_k$, the scale-dependent effective average action, from the microscopic action $\Gamma_\Lambda=S$ at the UV-cutoff scale $\Lambda$ to the macroscopic effective average action $\Gamma=\Gamma_0$, which encodes the effective equations of motion for the average fields, all fluctuation effects included. This is facilitated by the regulator $R_k$ which freezes out fluctuations at scales larger than the length scale $k^{-1}$. The boundary conditions of $\Gamma_k$ are enforced by requiring $R_\Lambda \sim \infty$ and $R_0=0$. $\Gamma_k$ contains all information about the statistics of the theory with fluctuations until the scale $k^{-1}$ incorporated. For example the inverse of $\Gamma_k^{(2)}$, the second order functional derivative of $\Gamma_k$, contains the correlation and response functions.

In general, Eq.~\eqref{eq:wetterich} can not be solved exactly and one has to resort to an approximation scheme, specified by an ansatz for the scale-dependent effective average action $\Gamma_k$ and the regulator $R_k$. Since Eq.~\eqref{eq:wetterich} does not hinge on the expansion of a small parameter, contrarily to the DRG formalism, these approximations are a priori nonperturbative.

For the regulator, we choose an algebraic cutoff \cite{morris_plb94}, which was previously used in polar active fluids \cite{jentsch_prr23}, except that it only acts on the wavevector component perpendicular to the collective direction of motion $\bq_\perp$. In Fourier space, it can be written as,
\begin{equation}
\label{cutoff2}
R_k(\tilde\bq,\tilde\bp) =\mu_{\bot,k} \frac{k^4}{q_\bot^2}\begin{pmatrix}
{ {\bf 0}} & { {\bf I}} &{ {\bf 0}} & { {\bf 0}}\\
{ {\bf I}} & { {\bf 0}} & { {\bf 0}} & { {\bf 0}}\\
{ {\bf 0}}  & { {\bf 0}}& 0 & 0\\
{ {\bf 0}} &{ {\bf 0}}& 0 & 0
\end{pmatrix}\tilde \delta_{qp} \ ,
\end{equation}
%
where {\bf I} denotes a $d$-dimensional identity matrix, $\tilde\bq=(\bq,\omega_q)$ and $\tilde\delta_{qp}=(2\pi)^{d+1}\delta^d(\bq+\bp)\delta(\omega_q+\omega_p)$. With this regulator choice all momentum integrals appearing in the trace of Eq.~\eqref{eq:wetterich} can be taken analytically \cite{SI}.

To confirm whether the results of Ref.~\cite{chen_njp18} remain valid in a nonperturbative setting, we choose an ansatz for $\Gamma_k$ that contains all terms present in the microscopic action, including the cubic coupling $b_k$ that has been neglected in Ref.~\cite{chen_njp18} and additionally two nonlinear momentum-dependant terms, characterized by $z_{k}^{(0)}$ and $z_{k}^{(1)}$,%
%
%
\begin{widetext}
\begin{align}
\label{eq:ansatz}
\Gamma_k[\bar\bv,\bv,\bar\cP, \cP] =& \int_{\tilde \br}\bigg\{ \bar\bv\cdot\bigg[\gamma_k \pp_t \bv +\lambda_k (\bv \cdot \vnab) \bv + \vnab \cP -\mu^\bot_k \vnab_\bot^2 \bv -\mu^x_k \partial_x^2 \bv +(a_k+b_k |\bv|^2)\bv \bigg]
 - D_k|\bar\bv|^2+ \bar\cP \nabla\cdot  \bv\\ \nonumber
&\hspace{0.8cm} -z_{k}^{(0)}\mathrm{Tr}\;\bar \bv \nabla_{\bot}
\cdot\bigg[\left(|\bv|^2 - v_{0,k}^2\right) \nabla_{\bot, j} \bv\bigg]
-z_{k}^{(1)}\bar \bv \cdot \partial_{x} \bigg[\left(|\bv|^2 - v_{0,k}^2\right) \partial_{x} \bv\bigg]\bigg\} \ ,
\end{align}
\end{widetext}
where $ v_{0,k}=\sqrt{|a_k|/b_k}$.
Our motivation for including these terms is to (a) check whether $b_k$ is actually an irrelevant coupling as claimed in Ref.~\cite{chen_njp18}, (b) try and shift the fixed point location and thus potentially change the value of the scaling exponents, as in the case of the critical $O(N)$ model and many other systems \cite{balog_prl19,depolsi_pre20}, and (c) introduce couplings that could create graphical corrections for the $q_x$ dependent part of the propagator, potentially breaking one of the hyperscaling relations found in the perturbative approach. In the perturbative calculation at one-loop level \cite{chen_njp18} these graphical correction are vanishing. From a perturbative viewpoint, the additional coupling $z_k^{(1)}$ incorporates higher order loop effects which could change this picture. Note that we could add up to seven additional momentum-dependent nonlinear terms of the same order, but only those included contribute to the self-energy of the Goldstone mode \cite{SI}.
Further, since the terms containing the pressure field and its response are linear they do not get renormalized \cite{canet_pre15,jentsch_prr23}. Therefore, we set their coefficients to unity. As derivatives in Eq.~\eqref{eq:ansatz} are split into contributions parallel and transverse to the $x$-direction, our ansatz seemingly breaks the rotational symmetry explicitly, however, all couplings can be identified with a fully symmetric ansatz \cite{SI}. Finally, nonperturbative contributions could also arise from the regulator choice: due to its dependence on $\mu_{\bot,k}$, the graphical correction of $\mu_{\bot,k}$ will be defined recursively, leading to flow equations that are nonpolynomial in the interaction terms.

The RG flow equations can now be deduced from Eq.~\eqref{eq:wetterich}, evaluated around the expectation value of the velocity $\bv(\bx,t)=\bv_0$ and in the comoving frame by setting external frequencies equal to $\omega = \lambda_k q_x v_0$ \cite{SI}. All but the Goldstone mode propagators are set to zero since they are of subleading order \footnote{Technically, there one contributing term, which we have included. See, \cite{SI}}. This is also justified a posteriori \cite{SI}.

Expressing the scale-dependent coefficients in Eq.~(\ref{eq:ansatz})
in dimensionless units (defined in \cite{SI} and denoted with an overbar here), the flow equations read
\begin{align}
\label{flow1}
\partial_l \gamma_k &= \partial_l D_k = 0 \ , \  \partial_l \mu_k^\bot = \eta_k^\bot \mu_k^\bot \ , \ \partial_l \mu_k^x = \eta_k^x \mu_k^x \\
\partial_l \bar\lambda_k &= \frac{1}{2}\left(4-d-\frac{5}{2}\eta^\bot_k-\frac{1}{2}\eta^x_k\right)\bar\lambda_k \ , \\
\partial_l \bar b_k &= \left(4-d-\frac{3}{2}\eta_k^\bot-\frac{1}{2}\eta^x_k\right)\bar b_k + f_b \ , \\
\label{flowl}
\partial_l \bar z_k^{(a)} &= \left(2-d-\frac{3-2a}{2}\eta_k^\bot-\frac{1+2a}{2}\eta_k^x\right)\bar z_k^{(a)} + f_z^{(a)} \ ,
\end{align}
where $l=-\log k/\Lambda$, and the detailed expressions for the $f$'s and $\eta$'s are given in Ref.~\cite{SI}. 

At a fixed point of the flow equations [Eqs.~\eqref{flow1}-\eqref{flowl}], the scaling dimension of the Goldstone modes can be extracted from the $k$-dependence of the equal-time correlation function [contained in $(\Gamma_{k=q}^{(2)})^{-1}$] \cite{SI,blaizot_pre06} (in Fourier space),
\begin{align}
\label{eqtime}
&C_T(\bq)\equiv\int d\omega \,  \langle\bu_T(\bq,\omega)\cdot\bu_T(-\bq,-\omega)\rangle \\
\nonumber
 & \approx\left. \,\frac{2D_k}{\gamma_k \mu_k k^2}  \int d\bar \omega\frac{\delta_{ij}-\hat x_i\hat x_j-\hat q_{\bot,i} \hat q_{\bot,j}}{|-\ii \bar\omega+\bar q_\bot^2+\bar q_x^2|^2}\right|_{k=q} \sim q^{\alpha_T}\ ,
\end{align}
with the dimensionless wavenumbers and frequencies,
\begin{equation}
\bar q_\bot = \frac{q_\bot}{k} \ , \ \bar q_x = \frac{q_x}{k}\sqrt{\frac{\mu_k^x}{\mu_k^\bot}} \ , \ \bar \omega_\bot = \frac{\omega_\bot \gamma_k}{\mu_k^\bot k^2} \ ,
\end{equation}
and $\alpha_T=\eta^\bot-2$. This exponent is related to the other exponents via $2\chi=-\alpha_T-d+1-\zeta$. 

Note that in our approximation scheme, $\mu_k^x$ does acquire a graphical correction \eqref{flow1}. If $\eta_k^x$ were to take a nonzero value at the fixed point, this would imply that the scaling exponents obtained in \cite{chen_njp18} receive graphical corrections and are thus not exact. 

The flow equations [Eqs.~\eqref{flow1}-\eqref{flowl}] can be integrated straightforwardly at different initial conditions to obtain the flow diagram, Fig.~\ref{fig:flow}a. Besides the trivial Gaussian FP (yellow pentagon), it contains 3 other nontrivial FPs (modulo the sign of $\bar \lambda_k$). 

{ {\bf FP 1:}} On the manifold $\bar b_k=0$, we find the fixed point (blue diamond) whose existence was assumed in Ref.~\cite{chen_njp18}. We have shown here explicitly that it exists (in $d=3$ the fixed point values are: $\bar\lambda_*=\pm 5.5$ and all other couplings vanishing) and confirm the scaling exponents that have previously been found, $\alpha_T=-2(d+1)/5$ \eqref{exact}. 

{ {\bf FP 2:}} On the other manifold, where $\bar \lambda_k=0$, we find the fixed point ($\bar b_*=8.6$ and all other couplings vanishing in $d=3$) associated to the Goldstone regime of the $O(N)$ model ($N=(d-1)$ here, since $\bv$ is a vector in realspace and one mode is removed by the incompressibility condition). At this fixed point, the scaling behavior of the Goldstone modes remains unmodified from the mean-field behavior $\alpha_T=-2$, however, the mode parallel to the flocking direction, $u_x=\bu\cdot \hat x$, becomes soft with a scaling dimension $\alpha_x=d-4$, which is different from mean-field theory, where one would expect this mode to have a finite correlation length \cite{patashinskii_ZhETF73, fisher_pra73,anishetty_ijmpa99,dupuis_pre11}. In the ERG formalism, this can again be seen from the equal time correlation [analogously defined as in Eq.~\eqref{eqtime}] \cite{dupuis_pre11},
\begin{align}
\nonumber
\label{longitudinal_scaling}
C_x(\bq) &\approx  \left. \,\frac{2D_k}{\gamma_k \mu_k k^2}  \int d \bar\omega\frac{1}{|-\ii \bar\omega+2\bar b_k \bar v_{0,k}^2+\bar q_\bot^2+\bar q_x^2|^2}\right|_{k=q} \\
&\xrightarrow{k\rightarrow 0} \left. \frac{D_k}{2\gamma_k b_k v_{0,k}^2}\right|_{k=q} \sim q^{\alpha_x} \ , 
\end{align}
so generally $\alpha_x=\partial_l \log(b_k)$ in the large $k$ limit, since $v_{0,k}$ approaches a fixed value in physical dimensions and $D_k$ and $\gamma_k$ do not renormalize. Therefore, close to the Gaussian fixed point (yellow pentagon, $l\lesssim10$ in Fig.~\ref{fig:flow}b and \ref{fig:flow}c) $\alpha_x=0$, indicating exponential decay of the correlation function due to a finite correlation length. However, at the Goldstone fixed point (green square), the dimensionless $\bar b_k$ takes a fixed point value such that $b_k$ vanishes asymptotically with 
\beq
\label{eq:alphax}
\alpha_x= d-4+\frac{3}{2}\eta^\bot+\frac{1}{2}\eta^x
\ .
\eeq
Since $\eta^\bot=\eta^x=0$ at this fixed point, we recover $\alpha_x=d-4$.

The longitudinal mode, parallel to the transverse momentum, $u_L = \hat \bq_\bot\cdot \bu$ is enslaved to the mode in the $x$-direction via the incompressibility condition $q_\bot u_L=-q_x u_x$, and therefore takes the same scaling dimension $\alpha_L=\alpha_x$, since there is no anisotropic scaling at this fixed point. The RG flow of the couplings and of the scaling dimensions close to this fixed point is shown in Fig.~\ref{fig:flow}b and \ref{fig:flow}c for values of $15\lesssim l\lesssim30$.

{ {\bf FP 3:}} Now we turn to the attractive fixed point (red circle) that describes generically the ordered phase of IPAF. In this {\it active Goldstone regime}, both $\bar\lambda_k$ and $\bar b_k$ take nonvanishing fixed point values and the higher order coupling $z_k^{(0)}$ is generated (in $d=3$: $\bar\lambda_*=\pm 7.1$, $\bar b_*=5.7$, $z_*^{(0)}=-1.4$ and all other couplings vanishing). The coupling $z_k^{(1)}$, however, which generates the anomalous dimension in the $x$-direction $\eta^x_k\sim z_k^{(1)}$, vanishes such that $\eta_k^x=0$. Therefore, the scaling behavior of the Goldstone mode remains unmodified compared to the $\bar b_k=0$ fixed point, providing strong evidence that the exponents described in Ref.~\cite{chen_njp18} are indeed exact with $\alpha_T=-2(d+1)/5$. 

Interestingly, as in the equilibrium Goldstone regime, the same argument for the fluctuations in the $x$-direction \eqref{longitudinal_scaling} applies, yielding again Eq.~\eqref{eq:alphax}.
At this fixed point, however, $\eta^\bot=2(4-d)/5$, hence, $\alpha_x= 2(d-4)/5$. Further, since the scaling at this fixed point is no longer isotropic, we find that the scaling dimension parallel to the transverse momentum differs from that in the $x$-direction: $\alpha_L=\alpha_x+\eta^x-\eta^\bot=4(d-4)/5$. Again, the RG flow of the couplings and of the scaling dimensions close to this fixed point is shown in Fig.~\ref{fig:flow}b and \ref{fig:flow}c for values of $l\gtrsim 35$. 

To recapitulate, in the active Goldstone regime, the fluctuations in the flocking direction ($u_x$) and longitudinal fluctuations ($u_L$) become long-ranged due to interactions with the Goldstone modes.

Having determined the values of these additional exponents not considered in Ref.~\cite{chen_njp18}, we can further check that they are consistent with the ``nonlinear-$\sigma$ model" picture. Namely, if we assume that 
\beq
\sqrt{|\bv_\perp|^2+ v_x^2} = {\rm constant}
\ , 
\eeq
then $u_x \sim |u_\perp|^2$. Namely, the exponent governing the spatial decay of the equal-time $u_x$-$u_x$ correlation, $\chi_x$, is exactly $2 \chi$. Hence, $\alpha_x = 2\alpha_T +d-1+\zeta$, which gives the expected value of $2(d-4)/5$. Having found $\alpha_x$, one can then use the incompressibility condition again to determine the scaling dimension of $u_L$.


{\it Summary \& Outlook.---}Our exact renormalization group analysis not only confirms the exact scaling exponents in 3D for incompressible polar active fluids (IPAF) first described in Ref.~\cite{chen_njp18}, it also uncovers many novel features of the active matter system. 
First, we demonstrate the existence of the nontrivial renormalization group (RG) fixed point (as opposed to being presumed in Ref.~\cite{chen_njp18}). Second, we obtain the actual RG flow diagram (\fig \ref{fig:flow}a) that i) demonstrates the relevance of the coefficient $b_k$ [associated with the nonlinear term $|\bv|^2 \bv$ in \eqref{eq:eom}], which was omitted in the analysis of  Ref.~\cite{chen_njp18} [see Eq.~(\ref{eq:uperp})], and yet whose presence does not modify the exact scaling exponents; and ii) connects IPAF to the thermal $O(N)$ model (when $\lambda =0$). Third, we uncover two novel exact scaling exponents that describe the scaling behaviors of $\bu_L$ and $u_x$. 

Our work provides convincing evidence that exact scaling exponents (and potentially other universal quantities, such as amplitude ratios)  can be obtained using RG methods at non-Gaussian RG fixed points. In particular, exact calculations seem possible for systems where the number of scaling exponents plus allowed graphical corrections is smaller than or equal to the number of relevant coefficients in the equations of motion, as in the case of IPAF considered here.  An immediate and important future direction is therefore to identify the precise criteria for exact RG calculations to be possible.

\bibliography{references}

\end{document}


\title{Supplemental material for Can exact scaling exponents be obtained using RG?\\
An illustration  in incompressible polar active fluids }
\author{Patrick Jentsch}
\email{p.jentsch20@imperial.ac.uk}
\address{Department of Bioengineering, Imperial College London, South Kensington Campus, London SW7 2AZ, U.K.}
\author{Chiu Fan Lee}
\email{c.lee@imperial.ac.uk}
\address{Department of Bioengineering, Imperial College London, South Kensington Campus, London SW7 2AZ, U.K.}
\date{\today}

	\begin{abstract}
	\end{abstract}
	
\maketitle

\section{Ansatz and linear analysis}

The ansatz for the effective action $\Gamma_k$ presented in the main text looks like it breaks the continuous rotational symmetry of the theory explicitly, since it contains couplings for terms with derivatives oriented in the flocking direction and perpendicular to it. It is however equivalent to the following ansatz that manifestly respects the rotational symmetries,
%
\begin{align}
\label{eq:ansatz}
\nonumber
\Gamma_k[\bar\bv,\bv,\bar\cP, \cP] = \int_{\tilde \br}&\left[  \bar\cP \nabla_i v_i - D_k \bar v_i \bar v_i +
 \bar v_i \cdot\left\{\gamma_k \pp_t v_i +\lambda_k (v_j \cdot \vnab_j) v_i + \vnab_i \cP +U_k^\prime(\kappa) v_i  \begin{array}{c}\; \\ \; \end{array}\right.\right. \\
&\left.\left. \begin{array}{c}\; \\ \; \end{array} -  \nabla_j \left( Z_k(\kappa)\nabla_j v_i \right)- \nabla_j  \left( Y_k(\kappa) v_j v_m \nabla_m v_i \right) \right\} \right]\ ,
\end{align}
where $\kappa=|\bv|^2/2$ and the term $U_k^\prime$ is defined as a derivative such that $U_k$ is the scale-dependent effective potential. The two-point functions associated to this ansatz, evaluated in a uniform background field ($\bar\bv=\bar\cP=\cP=0$, ${ \bv} = \bv_\unif $) are,
\begin{align}
\Gamma_{k,ij}^{(2,0,0,0)}[{0, \bv_\unif}, 0,0](\tilde \bq,\tilde \bp) = & -2D_k \delta_{ij} \tilde\delta_{qp} \ , \\
\Gamma_{k,ij}^{(1,1,0,0)}[{0, \bv_\unif},0,0](\tilde \bq,\tilde \bp) =&\left\{ \left[-\ii \gamma_k \omega_q + \ii \lambda_k v_\unif q_x +U_k^\prime(\kappa_\unif)+Z_k(\kappa_\unif) q^2+Y_k(\kappa_\unif)(\bv_{\unif}\cdot \bq)^2 \right]\delta_{ij}+ U_k^{\prime\prime}(\kappa_\unif)v_{\unif,i}v_{\unif,j}\right\}\tilde\delta_{qp}\ , \\
\Gamma_{k,i}^{(1,0,0,1)}[{0, \bv_\unif},0,0](\tilde \bq,\tilde \bp) = &\; \ii q_i\ ,\\
\Gamma_{k,i}^{(0,1,1,0)}[{0, \bv_\unif},0,0](\tilde \bq,\tilde \bp) = & -\ii q_i\ ,
\end{align}
where we have used the following identities,
\begin{equation}
\label{fcon}
\frac{\delta }{\delta v_{j}(\tilde \bq)} v_{i}(\tilde \br) = \delta_{ij} e^{\ii \omega_q t - \ii \bq \cdot \br } \ , \ \ \frac{\delta }{\delta \bar v_{j}(\tilde \bq)} \bar v_{i}(\tilde \br) = \delta_{ij} e^{\ii \omega_q t - \ii \bq \cdot \br } \ , \ \ \frac{\delta }{\delta \cP(\tilde \bq)} \cP(\tilde \br) = e^{\ii \omega_q t - \ii \bq \cdot \br } \ , \ \ \frac{\delta }{\delta \bar\cP(\tilde \bq)} \bar\cP(\tilde \br) = e^{\ii \omega_q t - \ii \bq \cdot \br } \ , 
\end{equation}
which also define our Fourier-Conventions.
Note that the only difference between the contribution in the $x$-direction and the isotropic term is generated by $U_k$, which will lead to the freezing out of the correlations in this direction. In general, we could write down 7 other terms at the same order in derivatives as the terms characterized by $Z_k$ and $Y_k$, however, since those would not be isotropic, their linear contributions would only contribute to the subleading $x$- or longitudinal modes or create nonlinear couplings to the same modes. Both effects would be vanishing in the small $k$ limit anyways.

By expanding the functions around the minimum of the potential, $\bv_{0,k}$, where $U_k^\prime(\kappa_{0,k})=0$ we can identify the couplings from the main-text as,
\begin{equation}
\label{couplings1}
a_k = 2\kappa_{0,k} U_k^{\prime\prime}(\kappa_{0,k})\ , \ \ \ b_k = U_k^{\prime\prime}(\kappa_{0,k})\ , \ \ \ \mu_k^\bot = Z_k(\kappa_{0,k})\ , \ \ \ \mu_k^x = Z_k(\kappa_{0,k})+2 Y_k(\kappa_{0,k})  \kappa_{0,k} \ ,
\end{equation}
\begin{equation}
\label{couplings3}
z_k^{(0)} = Z_k^\prime(\kappa_{0,k})\ , \ \ \ z_k^{(1)} = Z_k^\prime(\kappa_{0,k})+2Y_k(\kappa_{0,k}) \ .
\end{equation}
All couplings can therefore be obtained from the two-point function, such that all flow equations can be obtained from the second-order functional derivative of the Wetterich equation,
\begin{equation}
\label{wetterich2}
\partial_l \Gamma_k^{(2)} = \partial_{l^\prime}\left[-\frac{1}{2} \mathrm{Tr}\; \Gamma_k^{(4)} \cG_k + \mathrm{Tr}\; \Gamma_k^{(3)} \cG_k \Gamma_k^{(3)} \cG_k \right]_{k^\prime=k} \ ,
\end{equation}
with $\cG_k$ being the regulated propagator
\begin{equation}
\cG_k =\left( \Gamma_k^{(2)} +R_{k^\prime}\right)^{-1} \ ,
\end{equation}
and $k$ and $k^\prime$ are being treated as independent scale variables, such that $\partial_{l^\prime}$ only acts on the $k$-dependance of the regulator.
Defining,
\begin{equation}
G_{k,ij}^{-1}(\tilde \bq) = \frac{1}{VT}\Gamma_{k,ij}^{(1,1,0,0)}[0,\bv_\unif,0,0](\tilde \bq,-\tilde \bq) \ ,
\end{equation}
where $VT=(2\pi)^{d+1}\delta^{d+1}(0)$ is the spatiotemporal Volume, the unregulated propagator $\tilde\cG_k$ can be obtained by inverting the two-point function $\Gamma^{(2)}$,
\begin{align}
\tilde\cG_k& = \left(\Gamma^{(2)}\right)^{-1}=
\begin{pmatrix}
-2 D_k &  \bG_{k}^{-1}(\tilde \bq)  & 0 & \ii \bq \\
 (\bG_{k}^T)^{-1}(-\tilde \bq) & 0 & -\ii \bq & 0 \\
0 & \ii \bq^T & 0 & 0 \\
 -\ii \bq^T & 0 & 0 & 0 
\end{pmatrix}^{-1} \\
&
=\begin{pmatrix}
0 &  \bG^T_{k,\bot}(-\tilde \bq)  & 0 & \bP_l^T(-\tilde \bq) \\
 \bG_{k,\bot}(\tilde \bq)   & 2 D_k \bG_{k,\bot}(\tilde \bq) \cdot \bG^T_{k,\bot}(-\tilde \bq) & \bP_r(\tilde \bq) &  2D_k \bG_{k,\bot}(\tilde \bq)\cdot\bP_l^T(-\tilde \bq) \\
0 & \bP^T_r(-\tilde \bq) & 0 & \frac{-\ii \bP^T_r(-\tilde \bq)\cdot (\bG_{k}^T)^{-1}(-\tilde \bq)\cdot\bq}{q^2} \\
\bP_l(\tilde \bq) & 2D_k \bP_l(\tilde \bq) \cdot \bG_{k,\bot}^T(-\tilde \bq) &  \frac{\ii \bq^T\cdot \bG_{k}^{-1}(\tilde \bq)\cdot\bP_r(\tilde \bq)}{q^2} & 2D_k \bP_l(\tilde \bq) \cdot \bP^T_l(-\tilde \bq) 
\end{pmatrix}
\end{align}
where we have defined,
\begin{equation}
\bP_r(\tilde \bq)= \frac{-\ii \bq+\ii  \bG_{k,\bot}(\tilde \bq)\cdot  \bG_k^{-1}(\tilde \bq)\cdot \bq}{q^2}, \ \ \ \ \ \ \bP_l(\tilde \bq)=\frac{-\ii \bq^T+\ii \bq^T\cdot \bG_k^{-1}(\tilde \bq)\cdot \bG_{k,\bot}(\tilde \bq)}{q^2} \ ,
\end{equation}
and the transverse propagator,
\begin{equation}
\bG_{k,\bot} = \bG_{k,x_\bot} + \bG_{k,T} \ , 
\end{equation}
%
with,
\begin{align}
\label{propT}
\bG_{k,x_\bot}(\tilde \bq) = G_{k,x_\bot}(\tilde \bq)\bP_{x_\bot}(\bq) &= \frac{\bP_{x_\bot}(\bq)}{-\ii \gamma_k \omega_q+ a_k \frac{q_\bot^2}{q^2} + \ii \lambda_k v_0 q_x +\mu_k^\bot q_\bot^2+\mu_k^x q_x^2} \\
\label{propx}
\bG_{k,T}(\tilde \bq)=G_{k,T}(\tilde \bq)\bP_T(\bq) &= \frac{\bP_T(\bq)}{-\ii \gamma_k \omega_q+ \ii \lambda_k v_0 q_x +\mu_k^\bot q_\bot^2+\mu_k^x q_x^2} \ ,
\end{align}
and
\begin{align}
P_{x_\bot,ij}(\bq)&=\left(\frac{q_\bot}{q}\hat x_i-\frac{q_x}{q}\hat q_{\bot,i}\right)\left(\frac{q_\bot}{q}\hat x_j-\frac{q_x}{q}\hat q_{\bot,j}\right) \\
P_{T,ij}(\bq)&=\delta_{ij}-\hat x_i \hat x_j -\hat q_{\bot,i} \hat q_{\bot,j} 
\ ,
\end{align}
which are the projector parallel to the component of $\hat\bx$ that is perpendicular to $\bq$ and the projector transverse to both $\bq$ and $\hat\bx$ resepectively.
The contributions from the pressure term, produce precisely the desired effect that all velocity correlation and response functions (corresponding to $\bar\bv$ and $\bv$ entries) are transverse to the wavevector $\bq$. Most of the correlation functions involving the pressure $\cP$ or its response field $\bar\cP$ (corresponding to all other entries) are nonzero, but do not enter the RG calculation since there are no interaction terms involving the pressure. To represent the graphical corrections diagrammatically later, we also adopt a graphical notation for these two propagators,
\begin{align}
\label{grpropt}
\diagram{3} &= \bG_{k,T,ij}(\tilde \bq) \ , \\ 
\label{grpropx}
\diagram{4} &= \bG_{k,x_\bot,ij}(\tilde \bq) \ ,
\end{align}
and the noise term,
\begin{equation}
\diagram{5} = 2D_k \ .
\end{equation}
External lines never imply a propagator (except in Eqns.~\eqref{grpropt}, \eqref{grpropx} and \eqref{trees}).

Further, from the two-point function $\tilde \cG_k$, the correlation function $\bC_\bot$ can be obtained, given by the $(\bv,\bv)$ entry of $\tilde\cG_k$, i.e.,
\begin{equation}
\bC_\bot(\tilde \bq) = 2 D_k \bG_{k,\bot}(\tilde \bq) \cdot \bG^T_{k,\bot}(-\tilde \bq) \ .
\end{equation}
Note, that $q_x^2 \bC_{xx} = q_\bot^2 \bC_{q_\bot q_\bot}$, as expected from the incompressibility condition $q_x u_x+ q_\bot u_L = 0$.

Since the contribution of $\bG_{k,x_\bot}$ is massive, i.e.,  $a_k$ has a linear scaling dimension of $2$, it will freeze out during the RG-flow and no longer contribute in the large-scale limit. We can therefore make the approximation that 
\begin{equation}
\label{Gxsimple}
\bG_{k,x_\bot}= \bP_{x_\bot}(\bq) \frac{q^2}{a_k q_\bot^2} \ ,
\end{equation}
which will become exact in the limit $k\rightarrow 0$. In most diagrams one can even set $\bG_{k,x_\bot}=0$ directly, leading to the vanishing of that diagram. However, there is one diagram that needs to be treated more carefully, which is discussed below.

By shifting the frequency $\omega_q\rightarrow \omega_q+ q_x\lambda_k v_{0,k}/\gamma_k  $, the only other coupling with positive scaling dimension, i.e., $\lambda_k v_0$, can be removed from the propagator. It corresponds to a Galilei shift to the comoving frame. As we show later, this can be done for both internal as well as external frequencies for all graphical corrections appearing later.

\section{Regulator choice}

As the regulator, we choose an algebraic cutoff,
\begin{equation}
\label{cutoff2}
R_k(\tilde\bq,\tilde\bp) =\mu_{\bot,k} \frac{k^4}{q_\bot^2}\begin{pmatrix}
{ {\bf 0}} & { {\bf I}} &{ {\bf 0}} & { {\bf 0}}\\
{ {\bf I}} & { {\bf 0}} & { {\bf 0}} & { {\bf 0}}\\
{ {\bf 0}}  & { {\bf 0}}& 0 & 0\\
{ {\bf 0}} &{ {\bf 0}}& 0 & 0
\end{pmatrix}\tilde \delta_{qp} \ ,
\end{equation}
where,
\begin{equation}
r(y) = \frac{1}{y}
\end{equation}
and that acts only on transverse momenta. With this choice, all integrals appearing in the graphical corrections can be taken analytically and, since the regulator is frequency independent, causality is preserved naturally. Since the regulator modifies the the linear theory only by an isotropic term, the regulated propagators are obtained straightforwardly,
\begin{align}
\label{propT}
G_{k,x_\bot}(\tilde \bq) &= \frac{1}{-\ii \gamma_k \omega_q+ a_k \frac{q_\bot^2}{q^2} + \ii \lambda_k v_0 q_x +\mu_k^\bot q_\bot^2+\mu_k^x q_x^2+\mu_{k^\prime}^\bot {k^\prime}^2 r(q_\bot^2/{k^\prime}^2)} \\
\label{propx}
G_{k,T}(\tilde \bq) &= \frac{1}{-\ii \gamma_k \omega_q+ \ii \lambda_k v_0 q_x +\mu_k^\bot q_\bot^2+\mu_k^x q_x^2+\mu_{k^\prime}^\bot {k^\prime}^2 r(q_\bot^2/{k^\prime}^2)} \ .
\end{align}
The limit from Eq.~\eqref{Gxsimple}, which must be taken in dimensionless units, is still valid in the regulated theory.
Further, since the scale derivative of \eqref{cutoff2},
\begin{equation}
\partial_l R_k(\tilde\bq,\tilde\bp) =(\eta_k^\bot-4) \mu_{\bot,k} \frac{k^4}{q_\bot^2}\begin{pmatrix}
{ {\bf 0}} & { {\bf I}} &{ {\bf 0}} & { {\bf 0}}\\
{ {\bf I}} & { {\bf 0}} & { {\bf 0}} & { {\bf 0}}\\
{ {\bf 0}}  & { {\bf 0}}& 0 & 0\\
{ {\bf 0}} &{ {\bf 0}}& 0 & 0
\end{pmatrix}\tilde \delta_{qp}  \ ,
\end{equation}
depends on $\eta_k^\bot$, the flow equation for $\mu_k^\bot$ defines a recursion relation for $\eta_k^\bot$, the solution of which is nonpolynomial in at least some of the couplings. For instance, the nonperturbative nature of using such a regulator is enough to reveal the nonperturbative fixed point of the KPZ equation in the RG flow \cite{canet_a05}, even if the scaling exponents are not well described in this case. We therefore use this regulator to check whether the same effect plays a role for incompressible polar active fluids. 

We have also studied the following regulator with a sharp cutoff,
\begin{equation}
\label{cutoff1}
R_k(\tilde\bq,\tilde\bp) =-2D \left[ \Theta(|\bq_\bot|-k)-1\right] \begin{pmatrix}
 {\bf I} &{\bf 0} & {\bf 0} & {\bf 0} \\
{\bf 0} & {\bf 0}  & {\bf 0} & {\bf 0} \\
{\bf 0} & {\bf 0}  & { 0} & { 0} \\
{\bf 0} & {\bf 0}  & { 0} & { 0} 
\end{pmatrix}  \tilde \delta_{qp} \ ,
\end{equation}
which also allows for an analytic solution of the same integrals as with the algebraic regulator, however, we found that the resulting RG-flow becomes divergent in $d=3$. In larger dimensions, the scaling exponents presented in the main paper remained however unchanged. Since it has been shown previously, that sharp regulators can produce unreliable results in nontrivial FRG approximations \cite{morris_NPhB96}, we believe that the observed divergence at $d=3$ was an artifact of using this particular regulator.

\section{RG flow equations}

As outlined above, the flow equations for all couplings \eqref{couplings1}-\eqref{couplings3} can be obtained from Eq.~\eqref{wetterich2}, for which we still need to determine the vertices $\Gamma_k^{(3)}$ and $\Gamma_k^{(4)}$. The only nonvanishing components of these vertices are
\begin{align}
\Gamma^{(1,2,0,0)}_{ijk}(\tilde \bq,\tilde \bh,\tilde \bk)
    =&\left[
    -\ii \lambda_k (k_j \delta_{ik}+ h_k \delta_{ij}) 
    +b_k  v_{0,k} (\delta_{ij} \hat x_k+\delta_{jk} \hat x_i + \delta_{ki} \hat x_j ) \right. \\
    \nonumber
    &-z_k^{(0)}v_{0,k}(\delta_{ij}\hat x_k \bq_\bot\cdot\bh_\bot+\delta_{ik}\hat x_j \bq_\bot\cdot\bk_\bot) \\
    \nonumber
    &\left. -z_k^{(1)}v_{0,k}(\delta_{ij}\hat x_k q_x h_x+\delta_{ik}\hat x_j q_xk_x)
    \right]\tilde\delta_{qhk} \ , \\
\Gamma^{(1,3,0,0)}_{ijkl}(\tilde \bq,\tilde \bh,\tilde \bk,\tilde \bl)=&\left[b_k (\delta_{kl} \delta_{ij}+\delta_{jl}\delta_{ik} + \delta_{il} \delta_{kj} ) \right. \\
\nonumber
&-z_k^{(0)}(\delta_{ij}\delta_{kl}\bq_\bot\cdot\bh_\bot+\delta_{ik}\delta_{jl}\bq_\bot\cdot\bk_\bot+\delta_{il}\delta_{jk}\bq_\bot\cdot\bl_\bot) \\
\nonumber
&\left.-z_k^{(1)}(\delta_{ij}\delta_{kl}q_xh_x+\delta_{ik}\delta_{jl}q_xk_x+\delta_{il}\delta_{jk}q_xl_x)\right]\tilde\delta_{qhkl} \ .
\end{align}
Imposing momentum conservation at each vertex, we introduce their graphical notation,
\begin{align}
\diagram{6} &= \frac{1}{VT} \Gamma^{(1,2,0,0)}_{ijk}(\tilde \bq,-\tilde \bh,\tilde \bh-\tilde \bq) \ , \\
\diagram{7} &= \frac{1}{VT} \Gamma^{(1,3,0,0)}_{ijkl}(\tilde \bq,-\tilde \bh,-\tilde \bk,\tilde \bh+\tilde \bk-\tilde \bq)
\end{align}
where both propagators [Eqns.~\eqref{propT} and \eqref{propx}] can connect to any of the in- or outgoing lines.

The graphical corrections to the 2-point functions, 
\begin{equation}
\bF_v(q_x,\bq_\bot,\omega_q) =\frac{1}{VT}\left. \partial_l \Gamma_k^{(\bar \bv\bv)}\right|_{{\bar\bv=0,\bv=\hat \bx v_{0,k}, \bar\cP=0,\cP=0}} = \frac{1}{VT} \left.\frac{\delta^2\partial_l \Gamma_k}{\delta \bar \bv(\tilde \bq)\delta \bv(-\tilde \bq)}\right|_{{\bar\bv=0,\bv=\hat \bx v_{0,k}, \bar\cP=0,\cP=0}} \ ,
\end{equation}
\begin{equation}
\bF_D(q_x,\bq_\bot,\omega_q) =\frac{1}{VT}\left. \partial_l \Gamma_k^{(\bar \bv\bar \bv)}\right|_{{\bar\bv=0,\bv=\hat \bx v_{0,k}, \bar\cP=0,\cP=0}} = \frac{1}{VT} \left.\frac{\delta^2\partial_l \Gamma_k}{\delta \bar \bv(\tilde \bq)\delta \bar \bv(-\tilde \bq)}\right|_{{\bar\bv=0,\bv=\hat \bx v_{0,k}, \bar\cP=0,\cP=0}} \ ,
\end{equation}
can then be represented as,
\begin{align}
\label{fg}
\bF_g(\tilde \bq) &= \frac{1}{2}\diagram{9} - \diagram{10}  - \diagram{11} \ , \\
\label{fd}
\bF_D(0) &= 0 \ .
\end{align}
Here, we have already set to zero all diagrams that contain the $\bG_{k,x_\bot}$-propagator, except for the last diagram in Eq.~\eqref{fg} hinted at above. The reason is that the three-point vertices contain contrbutions proportional to $v_{0,k}$, which diverges in dimensionless units, such that the total diagram remains finite. Since the first diagram in  Eq.~\eqref{fg} is independent of frequencies, the tilde on the wavevectors has been omitted. The usual Feynman rules apply and the $\partial_l^\prime$ derivative needs to still be applied to each expression. Internal wavenumbers and frequencies are not written explicitly. 

Plugging in the expressions for each diagram, we find,
\begin{align}
\label{diag1}
\nonumber
\diagram{12}
=&2  D_k\partial_{l\prime} \int_{\tilde\bh}\left[ G_{k,T}(\tilde \bh) G_{k,T}(-\tilde \bh)\right] \\
&\hspace{-2.5cm}\times \left.\left\{b_k\left[(d-2)\delta_{ij}+2P_{T,ij}(\bh)\right]+(d-2)\delta_{ij}\left(z_k^{(0)}q_\bot^2+z_k^{(1)}q_x^2\right) \right\} \right|_{k^\prime=k}  \ , \\
\label{diag2}
\nonumber
\diagram{13} =&2D_k\partial_{l\prime}\int_{\tilde\bh}\left[ P_{T,mn}(\bh) G_{k,T}(\tilde \bh) G_{k,T}(-\tilde \bh)   P_{T,kl}(\bq-\bh)  G_{k,T}(\tilde \bq-\tilde \bh) \right] \\
\nonumber
&\hspace{-2.5cm}\times \left[\ii \lambda_k (q_m \delta_{ik}+h_k \delta_{im})+b_k  v_{0,k} \delta_{mk} \hat x_i  \right]   \\
&\hspace{-2.5cm}  \times \left.\left[\ii  \lambda_k (q_n \delta_{lj}-h_j \delta_{ln}) +b_k  v_{0,k} \delta_{ln} \hat x_j  + z_k^{(0)}v_{0,k}(\delta_{ln}\hat x_j [\bh_\bot-\bq_\bot]\cdot\bh_\bot) +z_k^{(1)}v_{0,k}(\delta_{ln}\hat x_j [h_x-q_x] h_x) \right]\right|_{k^\prime=k} \ , \\
\label{diag3}
\nonumber
\diagram{14} =&2D_k\partial_{l\prime}\int_{\tilde\bh}\left[ P_{T,mn}(\bh) G_{k,T}(\tilde \bh) G_{k,T}(-\tilde \bh)   P_{x_\bot,kl}(\bq-\bh)  G_{k,x_\bot}(\tilde \bq-\tilde \bh) \right] \\
    \nonumber
 &\hspace{-2.5cm}\times \left[\ii\lambda_k (q_m \delta_{ik}+h_k \delta_{im})  +b_k  v_{0,k} (\delta_{im} \hat x_k+\delta_{mk} \hat x_i  )+z_k^{(0)}v_0\delta_{im}\hat x_k \bq_\bot\cdot\bh_\bot  +z_k^{(1)}v_0\delta_{im}\hat x_k q_x h_x \right]  \\
&\hspace{-2.5cm}\times \left.\left[\ii\lambda_k (q_n \delta_{lj}-h_j \delta_{ln})+b_k  v_{0,k} (\delta_{ln} \hat x_j+\delta_{nj} \hat x_l  )+z_k^{(0)}v_{0,k}\delta_{ln}\hat x_j [\bh_\bot-\bq_\bot]\cdot\bh_\bot +z_k^{(1)}v_{0,k}\delta_{ln}\hat x_j [h_x-q_x] h_x \right]\right|_{k^\prime=k}    \ .
\end{align}
Projecting Eq.~\eqref{diag3} transverse to both $\hat\bx$ and $\bq$, plugging in Eq.~\eqref{Gxsimple}, using $a_k=2\kappa_{0,k}b_k=v_{0,k}^2b_k$ and taking the limit $v_{0,k}\rightarrow\infty$ (technically speaking, in these physical dimensions, $v_{0,k}$ remains finite and all other couplings become small, but the effect is the same), it simplifies to,
\begin{align}
\label{diag4}
 P_{T,im}(\bq)\diagram{15}P_{T,nj} =&2D_k\partial_{l\prime}\left.\int_{\tilde\bh}\left[ G_{k,T}(\tilde \bh) G_{k,T}(-\tilde \bh)    \right] \left\{b_k     P_{T,im}(\bq) P_{T,mn}(\bh)  P_{T,nj}(\bq)\right\}\right|_{k^\prime=k}\ ,
\end{align}
where some terms vanished due to antisymmetry with respect to reflection of the $\bh$ integration. From the form in Eq.~\eqref{diag4} one can see that it cancels out the term proportional to $P_{T,ij}(\bh)$ from Eq.~\eqref{diag1}, once it is projected transverse to both $\hat\bx$ and $\bq$ and equipped with the prefactors from Eq.~\eqref{fg}. This effect is related to the fact that if one were to solve the EOM for $u_x$ in the large scale-limit, it would remove the nonlinearity arising from the $b_k$ coupling (i.e. in the equilibrium limit, where $\lambda_k=0$, the Goldstone modes becomes effectively Gaussian). The fact that $b_k$ is still a coupling that plays a role in this RG-calculation is that this cancellation will only happen fully on the tree level. Our approach however is based on the effective action, i.e., the 1PI-generating functional, where tree-level effects are intentionally removed. In other words, the 1-PI 4-vertex is finite but will be cancelled out by tree diagrams when calculating, e.g., the 4-point correlation functions after the effective action has been determined, i.e., when $k=0$.
 
For example, in the following correlation function,
\begin{align}
\label{trees}
\nonumber
&\bigg\langle   v_{T,i}(\tilde \bq)\bar v_{T,j}(\tilde \bj)\bar v_{T,k}(\tilde \bk)\bar v_{T,l}(\tilde \bl) \bigg\rangle \\
\nonumber
&= -\diagram{20} + \diagram{21}+\diagram{22}+\diagram{23} \\
&= 0 \ ,
\end{align}
only the first diagram is a 1-PI contribution, which is nonzero. However, the three following diagrams cancel its contribution in the limit of small $k$, such that the correlation function vanishes. In Eq~\eqref{trees}, external lines do correspond to a propagator.

From the $\bF$'s the flow equations for all the couplings [Eqns.~\eqref{couplings1}-\eqref{couplings3}] can be projected:
\begin{align}
\label{project_f}
\partial_l \kappa_{0,k} &= -\frac{1}{b_k}\frac{1}{d-2} \left[ \mathrm{Tr}  \bP_T(\bq) \bF_v(q_x,\bq_\bot,\omega_q+q_x\lambda_k/\gamma_kv_{0,k}) \right]_{\tilde \bq=0}	\ ,\\ 
\partial_l \gamma_k &=	 \frac{\ii}{d-2} \left[ \frac{\partial}{\partial \omega_q} \mathrm{Tr} \bm\bP_T(\bq) \bF_g(q_x,\bq_\bot,\omega_q+q_x\lambda_k/\gamma_kv_{0,k}) \right]_{\tilde \bq=0}	\ ,\\ 
\label{project_l}
%
%
\partial_l \mu_k^\bot &= \frac{1}{2(d-2)}  \left[ \frac{\partial^2}{\partial q_\bot^2}  \mathrm{Tr} \bm\bP_T(\bq) \bF_g(q_x,\bq_\bot,\omega_q+q_x\lambda_k/\gamma_kv_{0,k}) \right]_{\tilde \bq=0}			\ ,\\ 
\partial_l \mu_k^x &=	 \frac{1}{2(d-2)}  \left[ \frac{\partial^2}{\partial q_x^2} \mathrm{Tr}\bm \bP_T(\bq) \bF_g(q_x,\bq_\bot,\omega_q+q_x\lambda_k/\gamma_kv_{0,k}) \right]_{\tilde \bq=0}	\ ,\\ 
\partial_l \lambda_k &=	 -\frac{\ii}{g_0(d-2)} \left[ \frac{\partial}{\partial q_x}  \mathrm{Tr}\bm\bP_T(\bq) \bF_g(q_x,\bq_\bot,\omega_q+q_x\lambda_k/\gamma_kv_{0,k}) \right]_{\tilde \bq=0}		\ ,\\
\partial_l  b_k &= \frac{1}{v_{0,k}}\frac{1}{d-2} \frac{\partial}{\partial v_{0,k}}\left[  \mathrm{Tr} \bm\bP_T(\bq) \bF_g(q_x,\bq_\bot,\omega_q+q_x\lambda_k/\gamma_kv_{0,k}) \right]_{\tilde \bq=0} \ ,\\ 
\partial_l z_k^{(0)} &=  \frac{1}{v_{0,k}}\frac{1}{2(d-2)} \frac{\partial}{\partial v_{0,k}} \left[ \frac{\partial^2}{\partial q_\bot^2}  \mathrm{Tr} \bm\bP_T(\bq) \bF_g(q_x,\bq_\bot,\omega_q+q_x\lambda_k/\gamma_kv_{0,k}) \right]_{\tilde \bq=0}			\ ,\\ 
\partial_l z_k^{(1)} &=	 \frac{1}{v_{0,k}} \frac{1}{2(d-2)} \frac{\partial}{\partial v_{0,k}} \left[ \frac{\partial^2}{\partial q_x^2} \mathrm{Tr}\bm \bP_T(\bq) \bF_g(q_x,\bq_\bot,\omega_q+q_x\lambda_k/\gamma_kv_{0,k}) \right]_{\tilde \bq=0}	\ , \\
\partial_l D_k &=	- \frac{1}{2(d-2)} \left[  \mathrm{Tr} \bm\bP_T(\bq) \bF_D(q_x,\bq_\bot,\omega_q+q_x\lambda_k/\gamma_kv_{0,k}) \right]_{\tilde \bq=0}		\ ,
\end{align}
where the Galilei transformation mentioned above is implemented on both the external and internal frequencies, $ \omega_q$ and $ \omega_h$, via shifting the argument of the $\bF$'s and a shift in the frequency integration respectively, to ensure comoving fluctuations are captured. Upon rescaling the couplings to dimensionless units,
\begin{equation}
\bar \kappa_{0,k} = \frac{\gamma_k\sqrt{\mu_k^\bot\mu_k^x}}{k^{d-2}D_kS^{d-1}} \kappa_{0,k} \ , \ \ \bar\lambda_k =  \sqrt{\frac{D_k S^{d-1}}{k^{4-d}\gamma_k \mu^{\bot \, 3}_k } \sqrt{\frac{\mu^\bot_k}{\mu^x_k}}} \lambda_k \ , \ \ \bar b_k =  \frac{ D_k S^{d-1}}{k^{4-d}\gamma_k\mu_k^{\bot\, 2}}  \sqrt{\frac{\mu^\bot_k}{\mu^x_k}}b_k \ ,
\end{equation}
\begin{equation}
\bar z_k^{(0)}  = \frac{ D_k S^{d-1}}{k^{2-d}\gamma_k\mu_k^{\bot\, 2}}  \sqrt{\frac{\mu^\bot_k}{\mu^x_k}} z_k^{(0)} \ , \ \ \bar z_k^{(1)} = \frac{ D_k S^{d-1}}{k^{2-d}\gamma_k\mu_k^\bot\mu_k^x}  \sqrt{\frac{\mu^\bot_k}{\mu^x_k}} z_k^{(1)} \ , 
\end{equation}
and introducing the anomalous dimensions,
\begin{equation}
\eta_k^\bot = \frac{\partial_l\mu_k^\bot}{\mu_k^\bot} \ , \ \ \eta_k^x = \frac{\partial_l\mu_k^x}{\mu_k^x} \ , \ \ \eta_k^\gamma = \frac{\partial_l\gamma_k}{\gamma_k} \ , \ \ \eta_k^D = \frac{\partial_lD_k}{D_k} \ ,
\end{equation}
the flow equations take the form,
\begin{align}
\partial_l \bar \kappa_{0,k} &= (d-2+\eta_k^\gamma-\eta_k^D+\frac{1}{2}\eta_k^\bot+\frac{1}{2}\eta_k^x) \bar \kappa_{0,k} + f_\kappa \ ,\\ 
\partial_l \bar \lambda_k &= \frac{1}{2}(4-d+\eta_k^D-\eta_k^\lambda-\frac{5}{2}\eta_k^\bot-\frac{1}{2}\eta_k^x)\bar \lambda_k\ ,\\ 
\partial_l \bar b_k &= (4-d+\eta_k^D-\eta_k^\lambda-\frac{3}{2}\eta_k^\bot-\frac{1}{2}\eta_k^x)\bar b_k + f_b\ ,\\ 
\label{flowz0}
\partial_l \bar z_k^{(0)} &=  (2-d+\eta_k^D-\eta_k^\lambda-\frac{3}{2}\eta_k^\bot-\frac{1}{2}\eta_k^x)z_k^{(0)}	+ f_z^{(0)}	\ ,\\ 
\partial_l \bar z_k^{(1)} &=	 (2-d+\eta_k^D-\eta_k^\lambda-\frac{1}{2}\eta_k^\bot-\frac{3}{2}\eta_k^x) z_k^{(1)}	+ f_z^{(1)}\ , 
\end{align}
where,
\begin{align}
\eta_k^\gamma &= 0 \ , \\
\eta_k^D &= 0 \ , \\
\label{recursive}
\eta_k^\bot &=  \frac{(d-2)(4-\eta_k^\bot)\Gamma(\frac{6-d}{4})\Gamma(\frac{d}{4})}{16\sqrt\pi}\bar z_k^{(0)} -\frac{(4-\eta_k^\bot)(1+(5-2d)d)\Gamma(\frac{4-d}{4})\Gamma(\frac{2+d}{4})}{32\sqrt\pi(d^2-1)}\bar\lambda_k^2 \ , \\
\eta_k^x &= \frac{(d-2)(4-\eta_k^\bot)\Gamma(\frac{6-d}{4})\Gamma(\frac{d}{4})}{16\sqrt\pi}\bar z_k^{(1)}  \ , \\
f_\kappa &= -\frac{(d-2)(4-\eta_k^\bot)\Gamma(\frac{6-d}{4})\Gamma(\frac{d}{4})}{16\sqrt\pi} \ , \\
f_b &= -\frac{(d-2)(4-\eta_k^\bot)\Gamma(\frac{6-d}{4})\Gamma(\frac{d}{4})}{64\sqrt\pi} \bar b_k \left(d z_k^{(0)}+2 z_k^{(1)}\right) + \frac{(d-2)(4-\eta_k^\bot)\Gamma(\frac{4-d}{4})\Gamma(\frac{2+d}{4})}{16\sqrt\pi} \bar b_k^2 \ , \\
\nonumber
f_z^{(0)} &=- \frac{(4-\eta_k^\bot)}{512(d^2-1)\sqrt{\pi}} \left\{ \left[ 8  (d-2)(d^2-1) \left(d \bar z_k^{(0)}+ 2\bar z_k^{(1)} \right)\bar z_k^{(0)}+(6d-d^2)(2d^2-5d-1)\bar\lambda_k^2\bar b_k \right] \Gamma\left(\frac{6-d}{4}\right) \Gamma\left(\frac{d}{4}\right) \right.\\
&\hspace{1cm}\left.+4\left[ 8 (d-2)(d^2-1) \bar b_k\bar z_k^{(0)}+(2d^2-5d-1)\left((2+d)\bar z_k^{(0)}+2\bar z_k^{(1)}\right)\bar\lambda_k^2\right] \Gamma\left(\frac{4-d}{4}\right) \Gamma\left(\frac{2+d}{4}\right)\right\}\ , \\
f_z^{(1)} &= -\frac{(d-2)(4-\eta_k^\bot)}{64\sqrt{\pi}} \bar z_k^{(1)} \left[ \left(d \bar z_k^{(0)}+ 2\bar z_k^{(1)} \right) \Gamma\left(\frac{6-d}{4}\right) \Gamma\left(\frac{d}{4}\right)+4\bar b_k \Gamma\left(\frac{4-d}{4}\right) \Gamma\left(\frac{2+d}{4}\right)\right] \ .
\end{align}

To solve the flow equations, Eq.~\eqref{recursive} is first solved algebraically for $\eta_k^\bot$ and then the flow equations are solved numerically by integrating from different initial conditions. If $\bar\beta_k=0$ and all other couplings nonzero initially, the system is attracted to one of the blue diamonds in Fig.~1a in the MT. For $\lambda_k=0$ and all other couplings nonzero initially, the system is attracted to the green square in Fig.~1a of the MT and for completely generic couplings the system is attracted to one of the red circles in Fig.~1a of the MT. 

The flow diagram, Fig.~1a of the MT, is projected from the 4-dimensional coupling space (after having plugged in the expressions for the $\eta$'s, and ignoring $\bar\kappa_0$, since all other flow equations no longer depend on it) in the following way: First, since $z_k^{(1)}$ is always attracted towards zero at every fixed point, it is set to zero. Second, $z_k^{(0)}$ is interpolated in such a way, that the location and stability of the fixed points is not impacted, i.e., we set,
\begin{equation}
\bar z_k^{(0)}= \bar z_k^{(0)*}\frac{\bar\lambda_k^2}{\bar\lambda_k^*{}^2} \frac{\bar b_k}{\bar b_k^*} \ ,
\end{equation}
where $\bar z_k^{(0)*}$, $\bar\lambda_k^*$ and $\bar b_k^*$ denote the values of $\bar z_k^{(0)}$, $\bar\lambda_k$ and $\bar b_k$ at the attractive fixed point (red circle in Fig.~1a of the MT), i.e., for $d=3$: $\bar z_k^{(0)*}=-1.38$, $\bar\lambda_k^*=7.13$ and $\bar b_k^*=5.73$. Since $\bar z_k^{(0)*} = 0$ at all the other fixed points, the locations and stabilities of all fixed points remains unchanged.

In $d=3$ this produces the following two-dimensional set of ordinary differential equations,
\begin{align}
\label{flowproj1}
\partial_l \bar\lambda_k&= \frac{128 \bar\lambda_k \sqrt{\pi}-72 \bar z_k^{(0)*}\Gamma^2\left(\frac{3}{4}\right)\frac{\bar\lambda_k^3}{\bar\lambda_k^*{}^2} \frac{\bar b_k}{\bar b_k^*} -9  \Gamma^2\left(\frac{5}{4}\right)\bar \lambda_k^3}{
256 \sqrt{\pi}+16 \bar z_k^{(0)*} \Gamma^2\left(\frac{3}{4}\right)\frac{\bar\lambda_k^2}{\bar\lambda_k^*{}^2} \frac{\bar b_k}{\bar b_k^*}+2 \Gamma^2\left(\frac{5}{4}\right)\bar \lambda_k^2
} \ , \\
\label{flowproj2}
\partial_l \bar b_k&= \frac{128\sqrt{\pi}\bar b_k-64\Gamma^2\left(\frac{3}{4}\right)\bar z_k^{(0)*}\Gamma^2\left(\frac{3}{4}\right)\frac{\bar\lambda_k^2}{\bar\lambda_k^*{}^2} \frac{\bar b_k^2}{\bar b_k^*}- \Gamma^2\left(\frac{5}{4}\right)(5\bar\lambda_k^2+32\bar b_k)\bar b_k}{128 \sqrt{\pi}+8 \bar z_k^{(0)*} \Gamma^2\left(\frac{3}{4}\right)\frac{\bar\lambda_k^2}{\bar\lambda_k^*{}^2} \frac{\bar b_k}{\bar b_k^*}+ \Gamma^2\left(\frac{5}{4}\right)\bar \lambda_k^2} \ ,
\end{align}
from which Fig.~1a of the MT is produced. We also produced a flow diagram by simply solving the flow equation for the fixed point solution of $\bar z_k^{(0)}$ in terms of $\bar\lambda_k$ and $\bar b_k$. While this method seems more elegant on paper, the resulting flow trajectories in the projected two-dimensional plane was much less accurate compared to the method used here. This is probably coincidental.

\bibliography{references}